# The polarization of the binary system Spica, and the reflection of light from stars

**Jeremy Bailey[1,*], Daniel V. Cotton[1], Lucyna Kedziora-Chudczer[1], Ain De Horta[2], Darren Maybour[2]**

[1]School of Physics, UNSW Sydney, NSW 2052, Australia.

[2]Western Sydney University, Locked Bag 1797, Penrith South DC, NSW 1797, Australia

*Correspondence to: j.bailey@unsw.edu.au

**Close binary systems often show linear polarization varying over the binary period, usually attributed to light scattered from electrons in circumstellar clouds[1,2,3]. One of the brightest close binary systems is Spica (Alpha Virginis) consisting of two B type stars orbiting with a period of just over 4 days. Past observations of Spica have shown low polarization with no evidence for variability[4,5,6]. Here we report new high-precision polarization observations of Spica that show variation with an amplitude ~200 parts-per-million (ppm). Using a new modelling approach we show that the phase-dependent polarization is primarily due to reflected light from the primary off the secondary and vice versa. The stars reflect only a few per-cent of the incident light, but the reflected light is very highly polarized. The polarization results show that the binary orbit is clockwise and the position angle of the line of nodes is 130.4 ± 6.8 degrees in agreement with Intensity Interferometer results[7]. We suggest that reflected light polarization may be much more important in binary systems than has previously been recognized and may be a way of detecting previously unrecognized close binaries.**

Observations of the linear polarization of the unresolved Spica binary system were obtained over the period from 2015 May 24 to 2018 Sep 2 with three different telescopes and three different versions of our high-precision ferroelectric liquid crystal (FLC) based polarimeters. The High Precision Polarimetric Instrument (HIPPI)[8] was used on the 3.9m Anglo-Australian Telescope (AAT) at Siding Spring Observatory from 2015 to 2017. Mini-HIPPI[9] was used on the 35 cm Schmidt-Cassegrain (Celestron 14) telescope at the UNSW Observatory in Sydney, and HIPPI-2 was used on both the 3.9 m AAT and on the 60 cm Ritchey-Chretien telescope at Western Sydney University's Penrith Observatory. Most of the observations were made with no filter giving a broad band from 350 – 700 nm with an effective wavelength of about 450 nm. We reported[10] in 2016 evidence for variable polarization in Spica, and the periodic nature of the variation was apparent in preliminary Mini-HIPPI observations obtained that same year[9].

In Fig. 1 we show the measured polarization data from all instruments plotted against orbital phase, with zero phase at periastron. The data are plotted as fractional Stokes parameters $Q/I$ and $U/I$ which are related to the degree of polarization $p$ and position angle $\theta$ (measured eastwards from north) through $Q/I = p \cos 2\theta$ and $U/I = p \sin 2\theta$. Phase-dependent polarization variability is clearly visible in the data from all instruments with two peaks and troughs in the orbital cycle. This form of variability is similar to that seen in a number of binary systems. There is also significant scatter, which exceeds the formal errors on the data and may indicate the presence of variability on other timescales. The failure to detect variability in past observations[6] can be attributed to the limited precision and poor phase sampling.

As already noted variable polarization in binary systems is usually attributed to circumstellar material. However in the case of u Her[11] and LZ Cep[12] it has been suggested that reflection of light between the two binary components is the main mechanism. Since Spica is a detached binary, and therefore not likely to show substantial mass transfer leading to gas streams, we decided to first model the polarization processes arising from the stellar photospheres themselves. Two main processes can be expected. Firstly the stars are tidally distorted as a result of the gravitational interaction and therefore non-spherical. This will lead to polarization analogous to that seen in rapidly-rotating stars[13]. Secondly polarization will arise from the reflection of light from each binary component off the photosphere of the other component.

By reflection, we mean true reflection, i.e. light that is scattered by the stellar atmosphere without being absorbed. It should be noted that the term "reflection effect" has been used since early last century in the context of eclipsing binary light-curves. However this effect, as normally modelled, is not true reflection. Instead as noted by Eddington[14] and Russell[15], the term is "inaccurate" and refers to absorption, heating and re-emission of radiation.

True reflection from stars has been little studied, and the amount and the polarization of light reflected from stellar atmospheres of different spectral types does not seem to have been previously determined, either from observation or modelling. We therefore did initial calculations using stellar atmosphere models from the Castelli & Kurucz solar abundance grid[16]. We used our version[13] of the SYNSPEC spectral synthesis code[17] that has been modified to include polarized radiative transfer using the VLIDORT code from RT Solutions[18]. We calculated the percentage reflection of light at normal incidence in the form of the radiance factor (*I/F*) – the reflected light as a fraction of that from a perfectly reflecting Lambert surface. In addition we calculated the polarization of the reflected light for a 90-degree reflection (45 degree angle of incidence, 45 degree angle of reflection). The polarization calculations included the polarizing effects of electron scattering (important in hot stars) as well as Rayleigh scattering from H and He atoms and $H_2$ molecules (important in cooler stars). These results are shown in Fig. 2. It can be seen that most stars are quite poor reflectors with a solar-type star reflecting less than 0.1% of the incident light. The light that is reflected is, however, very highly polarized. This is because scattering is predominantly single scattering. However, for hotter stars and low-gravity stars where electron scattering becomes a more important component of the opacity, the reflectance increases to a few per-cent, and the polarization drops a little due to the increase in multiple scattering.

To model the Spica binary system we use the Wilson-Devinney formulation[19] as extended[20] to include eccentric orbits and non-synchronous rotation. This allows us to determine the geometry of the tidally-distorted stars and the distribution of gravity and hence temperature across their surfaces. We create a set of ATLAS9 stellar-atmosphere models to cover the range of $T_{eff}$ and log $g$ values needed for each star. For each stellar atmosphere model we use SYNSPEC/VLIDORT as described above to solve the polarized radiative transfer equation for a range of geometries for both thermal emission and reflected light. We use a grid of regularly spaced "pixels" to cover the observer's view of the system for a specific orbital phase angle. Then for each pixel that overlaps one of the stars we determine the log $g$ value and geometrical parameters for that pixel and interpolate in our set of model calculations to obtain the intensity and polarization of direct and reflected light for that pixel (see Methods for further details).

The inputs required for the model are largely determined by previous photometric and spectroscopic studies of Spica[21,22,23], which have led to the orbital and stellar properties listed in Table 1. An example of the modelling process is shown in Fig. 3 where we show the gravity distribution across the stars, the intensity and polarization of the reflected light, and the intensity and polarization of the total light including reflection and emission. It can be seen that although the reflected light is a very small fraction of the total light, it is very highly polarized. Furthermore the pattern of polarization, for each star centered on the other star, leads to a large integrated polarization, whereas the polarization of the emitted light largely cancels when integrated over the stars. The integrated polarization from the model is shown as the solid lines on Fig. 1. This modelled polarization curve has been least-squares fitted to the data points using a three-parameter model, where the parameters are a rotation in position angle, to allow for the orientation of the orbit on the sky, and the polarization offsets in Q/I and U/I.

There are no adjustable parameters in the model that change the amplitude of the polarization curve. The amplitude is determined entirely by the adopted parameters taken from the literature as listed in Table 1. Thus the agreement of the modelled amplitude with that observed is a good test of our assumptions, and verifies our hypothesis that the observed phase dependent polarization variation can be understood entirely in terms of the stellar photospheres. The agreement of the data with the model is even better if only the higher-precision AAT data is used (see Supplementary Fig. 1). The lower two panels of Fig. 1 show the contributions of various processes to the total amplitude. Most of the polarization variation is due to reflected light from the two stars. Tidal distortion of the primary causes mostly a constant polarization due to the rotational flattening of the star, with only a small phase dependent part. The tidal distortion of the secondary was included in the model but the effect is negligible.

The Q/I and U/I offset values determined in our fit must be added to the modelled binary polarization to fit the observations. The most likely interpretation of these offset values is that they represent the interstellar polarization along the line of sight to Spica. The distance to Spica is 77 ± 4 pc according to its Hipparcos parallax[24] or 84 ± 4 pc from interferometry[7]. Interstellar polarization is typically small (~0.2 to 2 ppm/pc) within the Local Hot Bubble[10,25,26] but increases steeply at larger distances. The distance of Spica puts it near the edge of the Local Hot Bubble and therefore the observed offset value of ~250 ppm seems plausible as interstellar polarization at this distance. Measurements of the nearby star 69 Vir[5], which given its spectral type is likely to be intrinsically unpolarized[10], are of a similar magnitude and therefore consistent with this conclusion.

The third fitted parameter applies a rotation to the polarization position angles to match the model to the observations. This determines the position angle of the line of nodes of the orbit for which we obtain $\Omega = 130.4 \pm 6.8$ degrees (See Supplementary Fig. 2). The intensity interferometer data[7] on Spica also determined $\Omega$ as $131.6 \pm 2.1$ degrees. The good agreement between these values provides further confirmation for our interpretation of the polarization. The polarimetry also shows that position angle decreases with orbital phase implying a clockwise sense of rotation of the orbit (since position angles are measured anticlockwise from north), which also agrees with the intensity interferometer determination[7].

Several orbital parameters adopted for our modelling have uncertainties or discrepant values in different studies. To test how this impacts on our results we have run models with altered values for key parameters. We find that changes in eccentricity and inclination comparable with the uncertainties do not significantly affect our results. Changes in the argument of periastron $\omega$ have the effect of moving the model in phase and this changes the fitted values of $\Omega$ and the polarization offsets. The errors we have quoted on these parameters derive almost entirely from propagating the 7 degree error on $\omega$ in this way.

Table 1 also includes the geometric albedos for each star which were calculated from our model by adjusting the orbital elements $\omega$ and $i$ to give a face-on fully-illuminated view of each star and then calculating the geometric albedo from the fractional reflected flux. The values (3.61 % for the primary, 1.36 % for the secondary) are, to our knowledge, the first ever determinations of geometric albedos for stars. They are lower than those of planets in our Solar system and near the bottom of the range seen in hot Jupiter type exoplanets[27], which range from 3% to 40%.

Our results show that reflection is the main cause of phase dependent polarization in Spica, and similar effects are expected in other binary systems. The polarization will be largest if the stars are hot, close and have low gravity. In favourable circumstances the amplitude could be much larger than the ~200 ppm observed in Spica, and may account for some or all of the polarization that has previously been attributed to scattering from circumstellar material.

The frequency of stellar multiplicity varies strongly with mass and is high for massive stars but the figures are still uncertain due to bias and incompleteness[28]. Close binaries seen in face on orbits would go undetected by both spectroscopic and photometric methods, but could be detected by reflected-light polarization. In the case of a face-on circular orbit the degree of polarization would be constant but the position angle would vary with phase leading to a detectable signal.

**Acknowledgments:** The work was supported by the Australian Research Council through Discovery Project grants DP140100121 and DP160103231. We thank Robert Spurr of RT Solutions for providing the VLIDORT software.

**Author contributions:** All authors participated in the observations of Spica and commented on the manuscript. D.V.C. developed the data reduction code, carried out the data reductions, ran the ATLAS9 models and investigated Spica's interstellar polarization. J.B. developed the binary system and polarized radiative transfer codes, carried out the modelling, and prepared the manuscript.

**Competing interests:** The authors declare no competing financial interests.

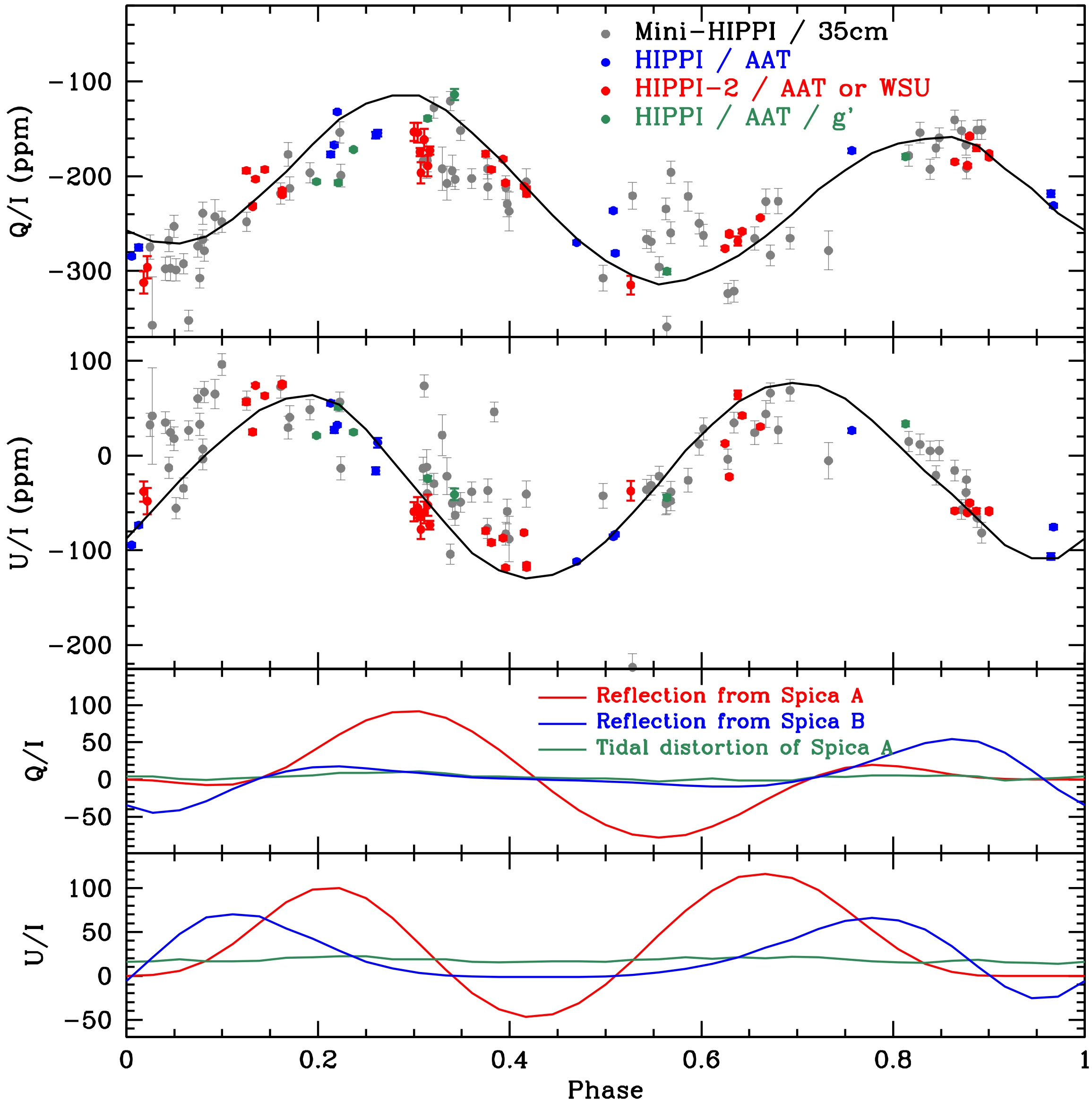

**Figure 1 | Polarization observations and modelling of Spica.** The observations were made with no filter except for the green points which were made in an SDSS g′ filter. Error bars are one-sigma internal errors derived from the statistics of the data. The black line is our modelled polarization curve for Spica for a wavelength of 450 nm. The lower panels of the plot show the contributions to the phase-dependent polarization deriving from reflected light from each of the stars (where star A is the primary and star B is the secondary) and from the tidal distortion of Spica-A. The contribution from tidal distortion of Spica-B is negligible.

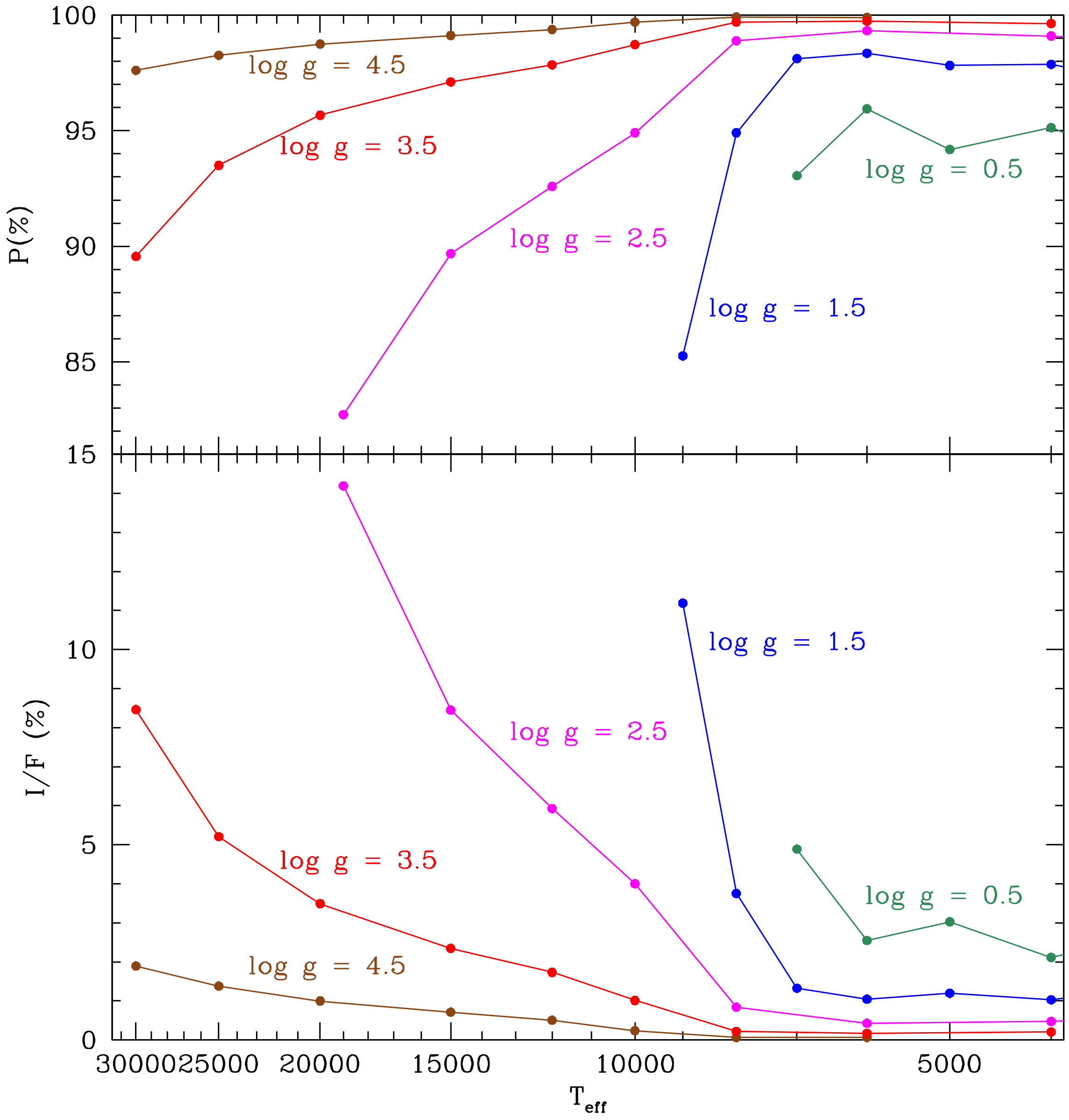

**Figure 2 | Reflected light and polarization for stellar atmosphere grid models.** These are calculations of the percentage reflectance (expressed as radiance factor I/F, which is the reflected light as a fraction of that from a perfect Lambert reflector) at normal incidence, and the polarization of the reflected light for a 90-degree reflection, at a wavelength of 440 nm. The calculations used the SYNSPEC/VLIDORT code applied to models from the solar abundance grid of Castelli & Kurucz[16].

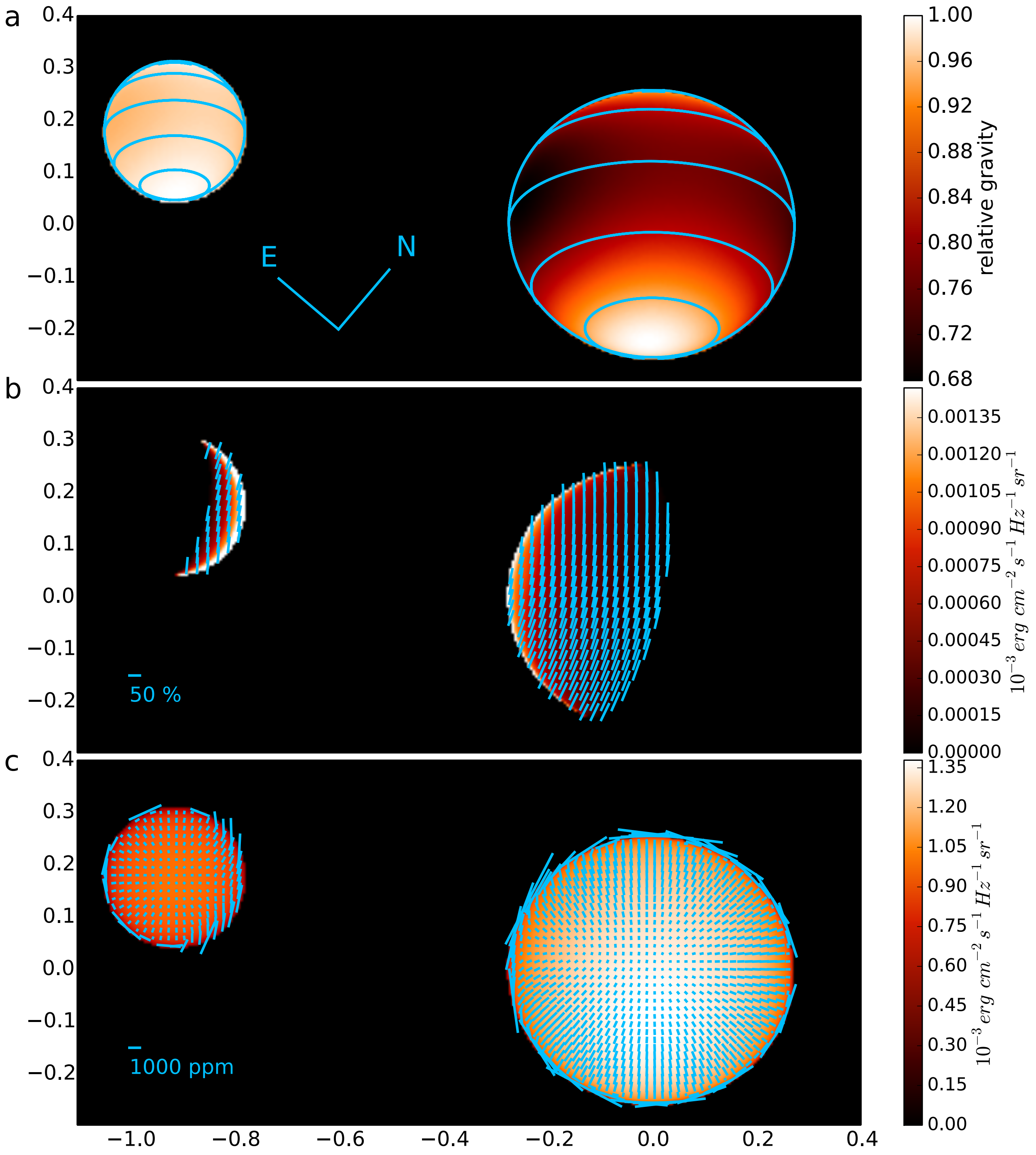

**Figure 3 | Polarization Modelling of Spica at Phase 0.222 (80 degrees). a.** The gravity distribution over the stellar surface, relative to that at the pole for each star. **b.** The intensity of reflected light only, over-plotted with vectors showing the polarization. **c**. The total intensity and polarization including emitted and reflected light. The pattern of the emitted polarization vectors is mostly radial except very near the limb where it switches to the tangential direction. A deviation from this pattern is apparent where reflected light is strong. All calculations are for a wavelength of 450 nm.

**Table 1 | Spica Model parameters.** A and B refer to the Primary and Secondary stars respectively.

| Name | Value | Reference | Notes |
|---|---|---|---|
| **Adopted Parameters** | | | |
| Epoch of Periastron | $T_0$ = MJD 40677.59 | 7 | |
| Period | $P$ = 4.0145898 days | 7 | |
| Argument of Periastron | $\omega$ = 288 ± 7 deg | 21 | From equation 1 of ref 19 for MJD 58000 |
| Inclination | $I$ = 63.1 ± 2.5 deg | 22 | |
| Eccentricity | $e$ = 0.133 ± 0.017 | 22 | |
| Mass ratio | $q$ = 0.6307 ± 0.023 | 22 | |
| Star A Potential | 4.67 | 22 | At periastron. As defined in ref 20. |
| Star B Potential | 6.08 | 22 | At periastron. As defined in ref 20. |
| Star A Rotation | $F$ = 2.0 | 22 | As defined in ref 20. |
| Star B Rotation | $F$ = 1.5 | 22 | As defined in ref 20. |
| Star A polar gravity | log $g$ = 3.78 | 23 | |
| Star B polar gravity | log $g$ = 4.16 | 23 | |
| Star A polar temperature | 24000 K | 23 | |
| Star B polar temperature | 19500 K | 23 | |
| **Derived Parameters** | | | |
| Position angle of line of nodes | $\Omega$ = 130.4 ± 6.8 deg | This work | Ref 7 gives 131.6 ± 2.1 |
| Offset in Q/I | –226 ± 19 ppm | This work | |
| Offset in U/I | –98 ± 6 ppm | This work | |
| Geometric Albedo Star A | $A_g$ (A) = 3.61 % | This work | |
| Geometric Albedo Star B | $A_g$ (B) = 1.36 % | This work | |

## Methods

**Observations and calibration.**

The observations of Spica used three telescopes and three instruments. The telescopes were the 35 cm Schmidt Cassegrain telescope at the UNSW observatory in Sydney, the 60 cm Ritchey Chretien telescope at Western Sydney University's (WSU) Penrith Observatory, and the 3.9 m Anglo-Australian Telescope at Siding Spring Observatory, New South Wales.

The instruments used were three different variants of our ferroelectric liquid crystal (FLC) based high-precision polarimeters. All instruments have in common the use of FLC modulators operating at 500 Hz, polarizing prism beam splitters and Hamamatsu H10720-210 photomultiplier tubes as detectors. The High Precision Polarimetric Instrument (HIPPI)[8] was used for observation on the AAT from 2014 to 2017. Mini-HIPPI[9] was used on the UNSW telescope, and our newest instrument HIPPI-2 was used for observations during 2018 on both the WSU telescope and the AAT.

HIPPI-2 is a new instrument that has replaced HIPPI. It is extremely versatile being small enough to mount on small telescopes, but also works effectively on large telescopes. HIPPI-2 uses the same modulator, Wollaston prism analyser, and detectors as HIPPI. However, the optical system has been redesigned for a slower input beam (f/16) and the simpler optics has led to significantly improved throughput. On telescopes with faster optics such as the AAT f/8 a relay lens can be used. HIPPI-2 uses a modulation and data acquisition scheme similar to that of Mini-HIPPI[9] with two stages of modulation, 500Hz from the FLC, and then a whole instrument rotation using a rotator built into the instrument. Compared with HIPPI, that used the telescope's own rotator, this makes observing faster and more efficient. The measured precision of polarimetry with HIPPI-2 on the AAT is comparable to or better than the 4.3 ppm precision initially obtained with HIPPI[8].

A bandpass model[8] is used to determine the effective wavelengths and modulation efficiency correction for each observation. The bandpass model takes account of the transmission of all the optical components in the different instrument configurations, and has recently been upgraded to take account of airmass as well. In terms of the Spica observations, for which we've assumed a B1V spectral type (representing the combined light of both components) without reddening, the transmission differences are largely restricted to wavelengths shortward of 400 nm. Two different FLCs were used with the three different instruments; a Micron Technologies (MT) unit was used exclusively for the Mini-HIPPI observations, whereas the same Boulder Non-Linear Systems (BNS) modulator was used for HIPPI and HIPPI-2 observations. The performance characteristics of the BNS unit drifted over time, and it was necessary to recalibrate it periodically using multi-band observations of high polarization standards. Tests carried out on the MT modulator showed its performance to be stable throughout our observations. We list the BNS performance characteristics for various eras in Supplementary Table 1.

The telescope tube and mirrors impart a small polarization known as the telescope polarization, or TP, which has to be subtracted from each measurement. We determine the TP by making repeat observations of a number of low polarization standard stars, a summary of which is given in Supplementary Table 2. On the AAT, so long as the mirrors are not realuminised or the instrument reconfigured, the TP is stable between runs, and successive data sets can be combined; otherwise each TP determination corresponds to a single run.

The polarization position angles (PA) also have to be calibrated. This is done with reference to polarized standard stars with known PAs, which are observed in either a clear or a green filter. A list of the standards used for each observing run is given in Supplementary Table 3. The errors in the standard PAs are of order 1 degree; consequently the PA calibration has a similar precision.

During July and August of 2018 the drift in modulator performance also resulted in a rotation of the PA at blue wavelengths. To correct for this observations made in clear were rotated to varying degrees based on their effective wavelength. The correction was based on a quadratic fit to g', 425 and 500 nm short pass band measurements of high polarization standards. The maximum PA correction applied to Spica observations in August was 2.0 degrees, in July it was less than 1 degree.

The full list of observations used in this study is given in Supplementary Table 4.

**Polarized radiative transfer.**

Polarized radiative transfer was carried out using the Vector Linearized Discrete Ordinate Radiative Transfer (VLIDORT) code[18]. This code has been widely used in Earth atmosphere applications but since it provides a completely general solution of the vector (i.e. polarized) radiative transfer equation, it is equally suited to applications in astronomy. In our past work we have applied it to the atmospheres of planets by incorporating it into our VSTAR planetary atmosphere code[29] and verified the results by reproducing past calculations of the polarization phase curves of Venus[30]. In addition we have applied it to stellar atmospheres[13] verifying the results against past calculations and successfully modelling the observed polarization in the rapidly rotating star Regulus[13]. The current work is an extension of that approach, which was based on integrating VLIDORT with the SYNSPEC spectral synthesis code[17].

The methods we have used previously[13] enable the calculation of the specific intensity and polarization emitted from a stellar atmosphere model as a function of wavelength and of the viewing angle ($\mu = \cos\theta$). The addition needed for this study is the calculation of the reflected light and its polarization. We make the simplifying assumptions that the external illumination is provided by an unpolarized point source, and that the structure of the stellar atmosphere is unchanged by the external illumination. It is then straightforward to calculate the reflected light intensity and polarization using VLIDORT. However, instead of requiring a single geometrical parameter ($\mu$), the result now depends on three parameters, the viewing zenith angle ($\mu$), the illuminating zenith angle ($\mu_0$) and the azimuthal angle between the two ($\phi - \phi_0$).

For the binary star models we used SYNSPEC/VLIDORT to calculate the specific intensity and polarization for each required stellar atmosphere model. Separate calculations were carried out for the emitted radiation and reflected radiation. In the emitted light case values were output for 21 $\mu$ values and can then be interpolated to give the result for any required viewing angle. In the reflected light case a coarse grid of geometrical parameters were used with 9 $\mu$ values, 9 $\mu_0$ values and 13 azimuths (at 30 degree intervals). Only 7 azimuths need to be calculated as the values above 180 degrees can be obtained by symmetry. Values for any required geometry are then obtained by 3D spline interpolation. The illuminating flux is set to one, and the results can later be scaled when that actual illuminating flux is known.

**Binary star modelling**

The binary star modelling is based on the extended Wilson-Devinney formalism[19,20]. The geometry of each star is described as an equipotential surface using an extended definition for the dimensionless potential $\Omega$ that allows for eccentric orbits and non-synchronous rotation[20]. We adopt a coordinate system in which the two star centres lie along the x-axis, the orbit is in the xy plane, and the centre of mass of the primary star is at the origin. The surface of the star has constant $\Omega$, and the vector ($d\Omega/dx$, $d\Omega/dy$, $d\Omega/dz$) is in the direction of the local normal to the surface, and has magnitude equal to the local effective gravity. We can also define the vector towards the secondary component, which, at a point $x, y, z$ on the surface of the primary is ($d-x, -y, -z$) where $d$ is the distance between the two stars, and define a third vector towards the observer, which depends on the orbital parameters and phase. From the relationships between these three vectors we can derive, for any point on the star's surface the geometrical parameters $\mu$, $\mu_0$ and $\phi-\phi_0$, described above, as well as the rotation angle $\xi$ needed to rotate polarization vectors from the frame of a local atmospheric model to the observers coordinate system.

The sizes of the two stars are defined by reference values for the dimensionless potential Ω given at periastron. For an eccentric orbit Ω is assumed to change with orbital phase in such a way that the volume of the star remains constant. We also need to specify the rotation parameters $F$ for each star, which is the ratio of the angular rotation rate to the synchronous rate. Our adopted values for these parameters are given in Table 1.

We adopt values for the polar temperature and polar gravity[21] of each star. At any point on the star we calculate the local effective temperature assuming von Zeipel's law, which for binary systems is found to be a reasonable approximation to more sophisticated recent models[31]. For each star we use a set of ATLAS9 stellar atmosphere models with a range of $T_{eff}$ and log $g$ values sufficient to cover the gravity variation over each star.

Following our previous approach[13] we adopt a uniform pixel grid with a pixel spacing of 0.0025 times the semi-major axis, in the observer's coordinate system, for a given orbital phase. For each pixel that overlaps a point on one of the stars we determine the local effective gravity and geometric parameters for that point, as well as the distance of the point from the other star. We interpolate in the set of model atmospheres, first in gravity, and then in geometry, to obtain the intensity and polarization of emitted light and reflected light for that pixel. We can use the data to plot images with overlaid polarization vectors as in Figure 3, or we can sum all the pixels to obtain the integrated polarization for each phase, to produce the modeled phase curves used in Fig. 1.

**Uncertainties in orbital parameters**

The binary parameters listed in Table 1 have uncertainties associated with them, and there are in some cases significant differences in values from different sources. In particular for the eccentricity we adopt[22] the value of $e = 0.133 \pm 0.017$ listed in Table 1, but an alternate source[21] gives a value as low as $0.108 \pm 0.014$. To test the effect on our results we have done additional model calculations with the values of $\omega$, $i$ and $e$ changed from their nominal values. The results are shown in Supplementary Fig. 3.

For the argument of periastron $\omega$, the nominal value of this time dependent (due to apsidal motion) parameter for the mid point of our observations is 288 degrees, but with an uncertainty of ±7 degrees[19]. We therefore ran models for the values of $\omega = 281$ and 295 degrees. We found that a good fit to the data could still be obtained, provided the three-parameter model fit was redone for the new models, since there was a strong correlation between the adopted value of $\omega$ and the fitted value of Ω and the polarization offsets. The uncertainties on the fitted parameters listed in Table 1 derive from this source.

Changing the eccentricity $e$ to the value of 0.108, as described above, leads to only a small change in the resulting phase curves, which are still a good fit to the observations. Changing the inclination $i$ by an amount comparable with the uncertainties also has only a small effect, and in both these cases there was no need to redo the three-parameter model fit. These results show that our conclusions are largely independent of uncertainties in the orbital parameters.

**Wavelength dependence of reflected light polarization**

The detailed phase-dependent modeling of the Spica binary system was carried out at a wavelength of 450 nm, approximately the effective wavelength of our observations. However, while it is much more computationally demanding, our modelling techniques allow calculation of the full wavelength dependent polarization. In Supplementary Fig. 4 we show the polarization of the reflected light from the primary, as a fraction of the total light from the system, for phase 0.222. This is the same phase shown in Fig. 3, and is about where the polarization due to reflected light reaches its maximum. It can be seen that the polarization of the reflected light increases to the blue up to about 390 nm and then falls at shorter wavelengths. The flux distribution of the binary system, from the same model is also shown, together with the instrumental and Earth atmosphere response for our various instrument and filter combinations.

The plot shows that the adoption of 450 nm for our modelling is a reasonable approximation. It does however, also show that small changes in the wavelength response, due for example to airmass differences, could account for some of the scatter seen in our observations.

**Simplifications**

We note that our current approach involves a number of simplifications in the modelling. While the geometry of the stars are modeled in detail when considering the reflection and emission, the illuminating star is treated as a point source in each case. Furthermore the atmosphere of the reflecting star is assumed to be unchanged in structure by the external illumination. The latter in particular is known to be incorrect and means that we are ignoring the heating, which is accounted for in the classical treatment of the misnamed "reflection effect". This effect is known to be important in correctly modelling binary light curves, but we believe this simplification has only a small effect on the resulting polarization, as it can be seen from Fig. 2 that a small change in temperature has only a very minor effect on the reflectance and polarization.

**Data availability:** All processed data generated during this study is included in this published article (and its supplementary information files), the raw data files are available upon reasonable request. All other data analysed in this work comes from public repositories, where this is the case the origin of the data is indicated in the text.

# Supplementary Information

## Supplementary References

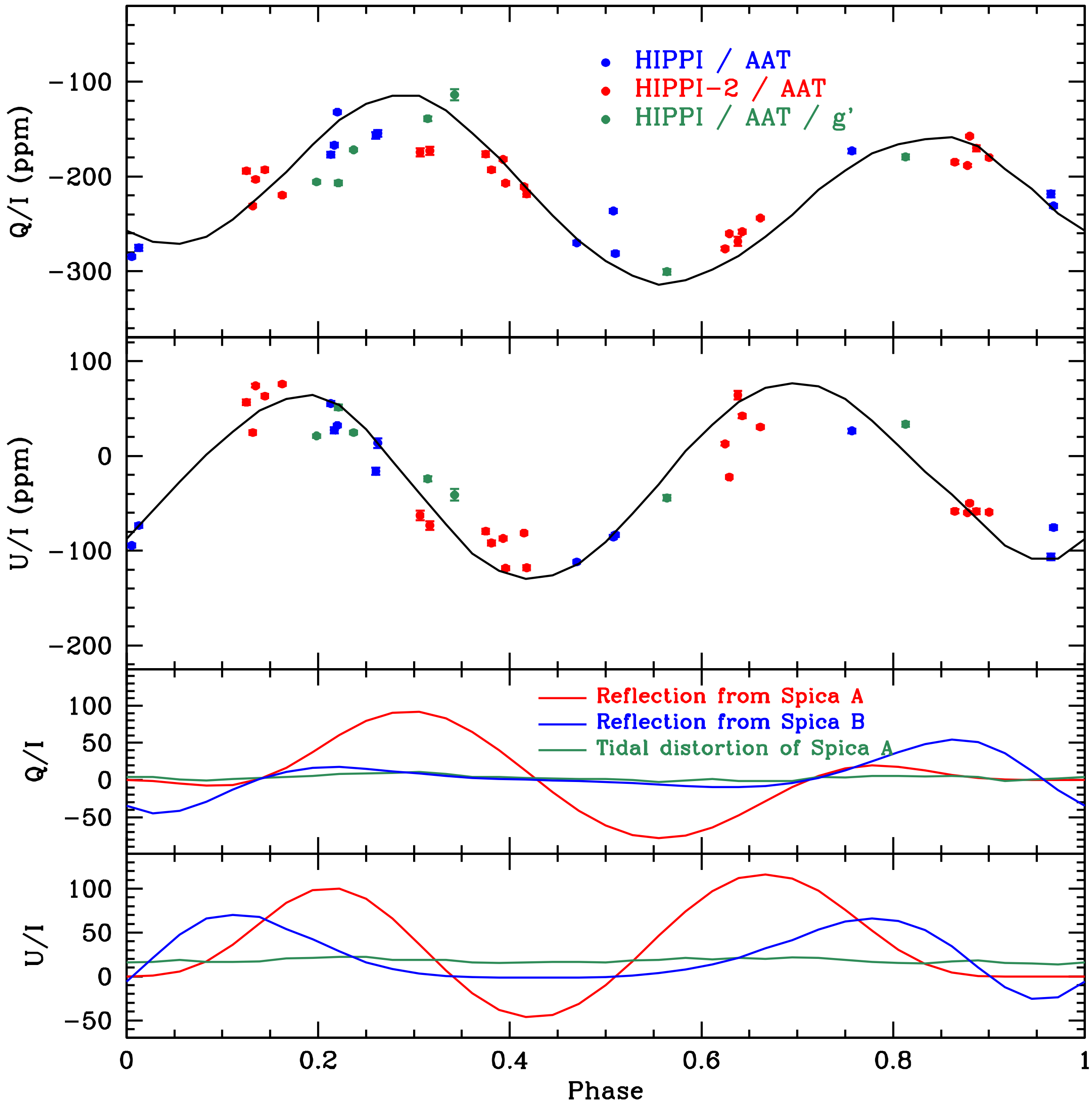

**Supplementary Figure 1 | AAT Observations**. As Fig. 1 but showing only the AAT observations which have higher precision than those on the smaller telescopes. It can be seen that the AAT observations show better agreement with the model curve.

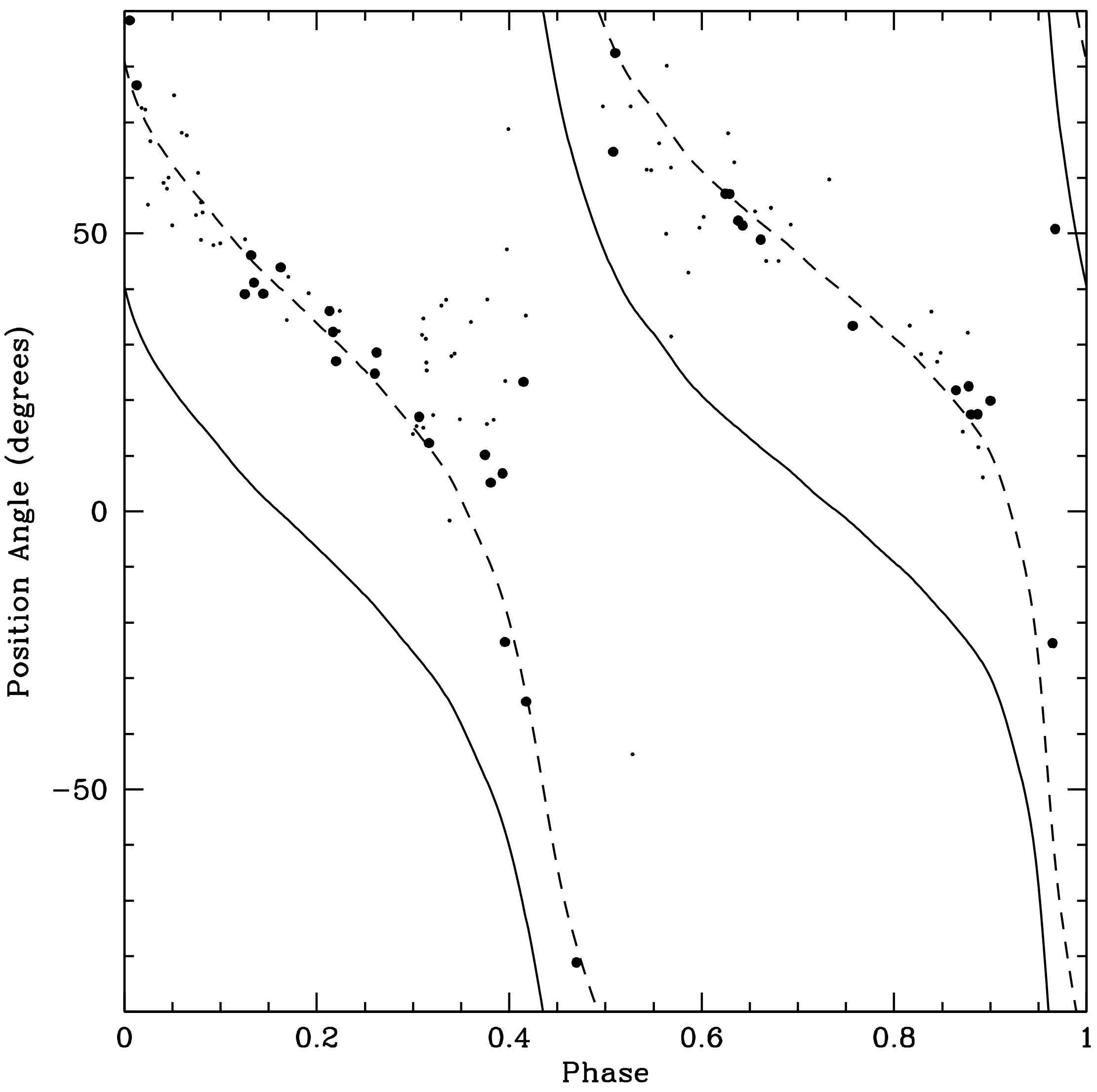

**Supplementary Figure 2 | Position Angle Variations.** The points show the observed polarization position angle of Spica after correcting for the zero point offsets listed in Table 1. Larger dots are AAT observations. The position angle decreases with phase implying a clockwise sense of orbital motion. The solid line is the modelled position angle – where the model is calculated with the line of nodes along the x-axis corresponding to Ω = 90 degrees. The dashed line shows the best fit, which is obtained by rotating the model by +40.4 degrees in position angle giving an actual position angle of the line of nodes of Ω = 130.4 degrees.

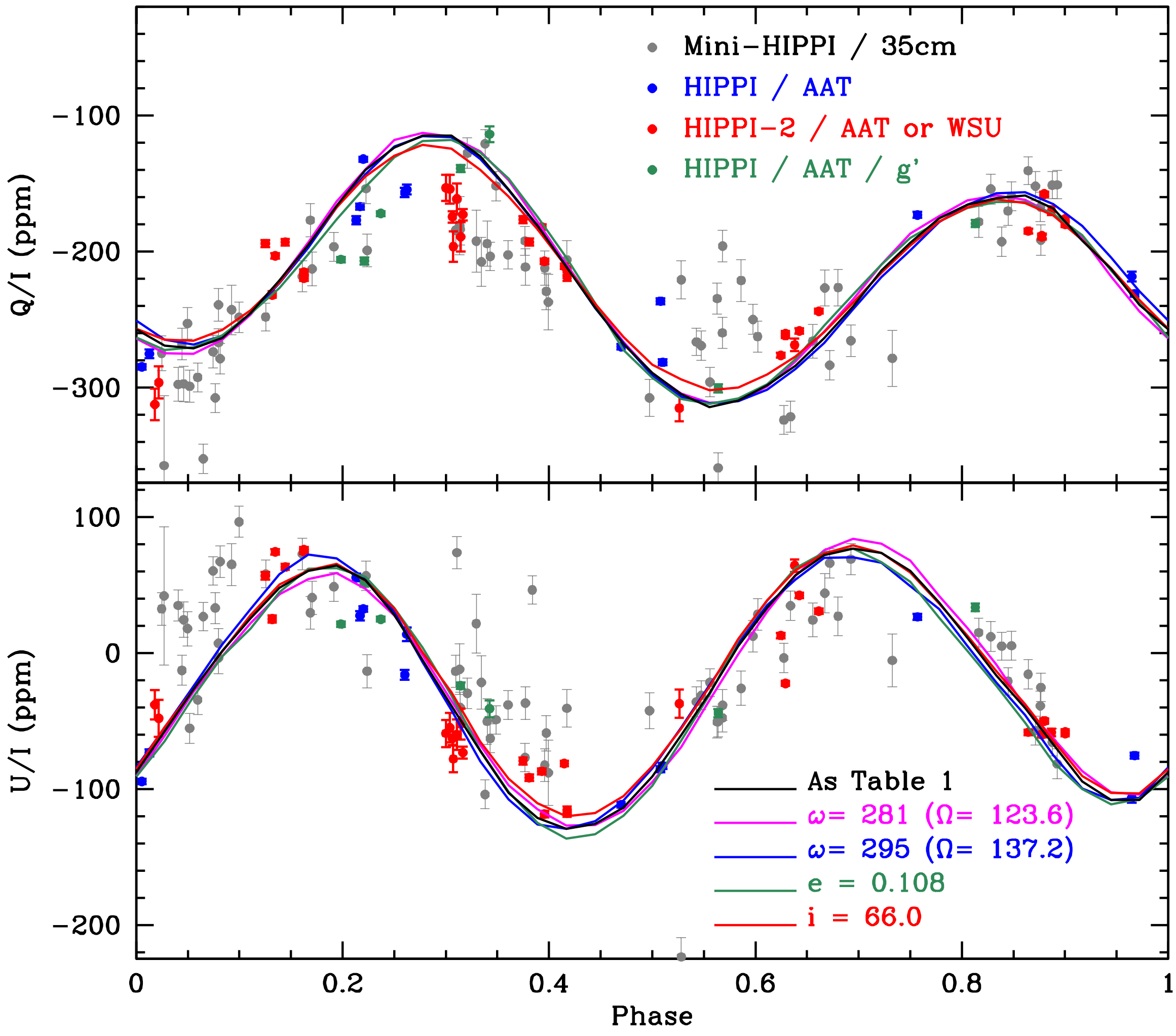

**Supplementary Figure 3 | Effect of Orbital Parameters.** The linear polarization data points are as for Fig. 1. The nominal model fit based on the parameters given in Table 1 is shown as the black line. Other curves show the result of adjusting the orbital parameters. The two values for the argument of periastron $\omega$ are the $\pm 7$ degree points given by the uncertainty[21] in the apsidal motion. These values result in a reasonable fit to the data provided the values of the polarization offsets and $\Omega$ are redetermined. The effects of changing eccentricity and inclination to more extreme values[21] are also shown. Changes comparable with the spread of values in the literature do not significantly change the results.

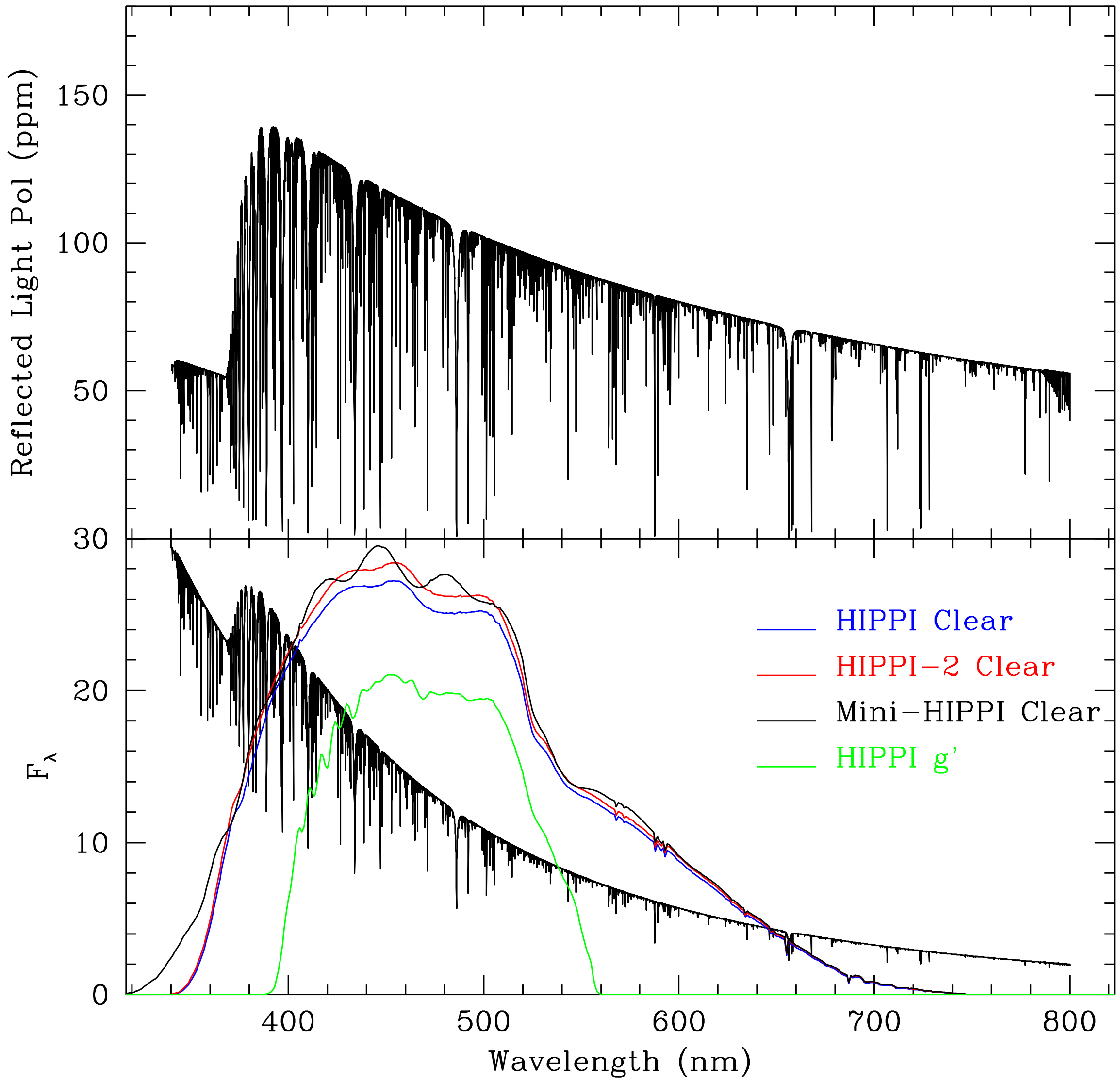

**Supplementary Figure 4 | Wavelength Dependence.** Modelled wavelength dependence of the reflected light polarization from the primary as a fraction of the total light from the system. The phase is 0.222, the same as that shown in Fig. 3. The lower panel shows the total flux for the same model, and the instrument plus atmosphere (for airmass 1) response for the various instrument/filter combinations.

**Supplementary Table 1 | Modulator Performance Characteristics.**

| Mod[a] | Era[b] | Run(s) | $\lambda_0$[c] | Cd[d] |
|---|---|---|---|---|
| BNS | 1 | 2015MAY, 2015JUN | 494.8 | 1.738E7 |
| BNS | 2 | 2016FEB, 2016JUN, 2017JUN, 2017AUG | 506.3 | 1.758E7 |
| BNS | 3 | 2018FEBB, 2018FEBC, 2018FEBD, 2018MAR, 2018MAY | 513.4 | 2.319E7 |
| BNS | 4 | 2018JUL | 520.6 | 2.127E7 |
| BNS | 5 | 2018AUG (HJD: 58346.37939 – 58352.39406) | 546.8 | 2.213E7 |
| BNS | 6 | 2018AUG (HJD: 58357.38917) | 562.7 | 2.319E7 |
| BNS | 7 | 2018AUG (HJD: 58359.37215 – 58363.36297) | 595.4 | 1.615E7 |
| MT | | M2016MAY, M2016JUN, M2016JUL, M2016OCT, M2018JAN | 505.0 | 1.750E7 |

## Notes:

[a] Mod indicates either the Boulder Nonlinear Systems (BNS) or Micron Technologies (MT) modulator. Data for the MT modulator comes from ref [8].

[b] The performance era of the BNS modulator.

[c] The wavelength in nm of peak modulation efficiency.

[d] C is a parameter describing the dispersion in birefringence of the FLC material and d is the thickness of the FLC layer; these two terms are treated as one constant (see ref [8]).

**Supplementary Table 2 | Telescope Polarization Measurements.**

| Run(s) | Configuration | | Adopted TP[c] | | Standards Observed[d] | | | | | | |
|---|---|---|---|---|---|---|---|---|---|---|---|
| | Tel[a]/ Instr | Fil[b] | Q/I (ppm) | U/I (ppm) | HD 2151 | HD 10700 | HD 49815 | HD 102647 | HD 102870 | HD 128620J | HD 140573 |
| 2015MAY, 2015JUN | AAT/ HIPPI | g’ | −35.6 ± 1.3 | 4.9 ± 1.3 | | | 2 | 1 | | | 4 |
| 2016FEB, 2016JUN | AAT/ HIPPI | g’ | −20.4 ± 1.8 | 4.2 ± 1.9 | 2 | | 4 | | 3 | | |
| 2017JUN, 2017AUG | AAT/ HIPPI | Cl | −14.8 ± 1.2 | 0.9 ± 1.2 | 6 | | 5 | 3 | 2 | | |
| 2018FEBB | AAT*/ HIPPI2 | Cl | −189.8 ± 1.1 | 8.8 ± 1.0 | | | 2 | 1 | | | |
| 2018FEBC, 2018FEBD | AAT*/ HIPPI2 | Cl | −178.7 ± 0.8 | 14.2 ± 0.8 | | | 5 | 3 | 1 | | |
| 2018MAR | AAT/ HIPPI2 | Cl | 130.3 ± 0.7 | 5.2 ± 0.7 | | | 4 | 4 | 3 | | 5 |
| 2018MAY | WSU/ HIPPI2 | Cl | −26.5 ± 2.0 | −2.8 ± 2.0 | | | 6 | 5 | | | |
| 2018JUL, 2018AUG | AAT/ HIPPI2 | Cl | −9.9 ± 1.0 | 3.7 ± 0.9 | 3 | 3 | 2 | 2 | 2 | | 2 |
| M2016MAY | UNSW/ MHIPPI | Cl | −93.8 ± 4.0 | −0.3 ± 4.0 | | | 4 | | | | |
| M2016JUN | UNSW/ MHIPPI | Cl | −61.9 ± 4.1 | −7.3 ± 4.3 | | | 5 | 1 | | | |
| M2016JUL, M2016OCT | UNSW/ MHIPPI | Cl | −69.4 ± 3.0 | −9.5 ± 3.0 | 4 | | 14 | 6 | 1 | | 1 |
| M2018JAN | UNSW/ MHIPPI | Cl | −58.0 ± 2.0 | 16.3 ± 2.0 | | | 9 | | | 3 | |

**Notes:**

[a] Telescopes are: UNSW, 35 cm Schmidt Cassegrain telescope at UNSW Observatory, Sydney; WSU, Western Sydney University’s 60 cm Ritchey Chretien Telescope at Penrith Observatory; AAT, 3.9 m Anglo-Australian Telescope.
[b] Cl denoting clear (no filter) or SDSS g’ filter.
[c] The adopted TP is the mean of all low polarization standard observations during the run(s).
[d] The observed standards have been shown previously to have very low polarizations, ~5 ppm or less: HD 2151[10], HD 10700[26] HD 49815[10], HD 102647[32], HD 102870[32], HD 140573[32]. On small telescopes we have also used HD 128620J, which has a measured polarization magnitude of ~10 ppm[9], owing to its brightness.
* Indicates AAT f/15, all other AAT observations are f/8.

**Supplementary Table 3 | PA Calibration Observations.**

| Run | ΔPA[a,b] (degrees) | Standards Observed | | | | | | | | |
|---|---|---|---|---|---|---|---|---|---|---|
| | | HD 80558 | HD 84810 | HD 111613 | HD 147084 | HD 149757 | HD 154445 | HD 160529 | HD 187929 | HD 203532 |
| 2015MAY | 0.2 | | | | 4 | | 1 | | | |
| 2015JUN | 0.1 | | | | 1 | | 1 | | | |
| 2016FEB | 0.3 | 1 | | | 1 | | | | | |
| 2016JUN | 0.2 | | | | 1 | | 1 | | | |
| 2017JUN | 1.1 | | | | 1 | | 1 | 1 | | |
| 2017AUG | 0.5 | | | | 1 | | 1 | | 1 | |
| 2018FEBB | – | 1 | | | | | | | | |
| 2018FEBC | 0.1 | 2 | | | | | | | | |
| 2018FEBD | – | 1 | | | | | | | | |
| 2018MAR | 0.3 | 1 | | 1 | 1 | | | | | |
| 2018MAY | 0.1 | | | | 2 | | | | | |
| 2018JUL[c] | 1.6 | | | | 1 | | 1 | 1 | 1 | 1* |
| 2018AUG[c] | 0.9 | | | | 3 | | | 3 | 5 | |
| M2016MAY | 0.1 | | 4 | | | | | | | |
| M2016JUN | 0.4 | | 2 | | 1 | | | | | |
| M2016JUL | 0.4 | | 1 | | 1 | | 1 | 1 | 1 | |
| M2016OCT | 0.1 | | 2 | | 1 | | | | | |
| M2018JAN | 0.5 | | | | 3 | 2 | | | | |

**Notes:**

[a] The standard deviation in PA from multiple measurements, using g' and clear filters only.
[b] The position angle is calibrated against the position angles of polarized standards known from the literature: HD 80558[32], HD 84810[33,34], HD 111613[33], HD 147084[35,36], HD 149757[37], HD 154445 [33,34,36], HD 160529[33,34], HD 187929[33], HD 203532[38]. The typical error in the literature determinations is of order a degree.

[c] A PA correction was applied to clear observations during these runs based on the effective wavelength ($\lambda$) according to:

July 2018: For $\lambda > 462$ nm, DPA = $-0.00460905\,\lambda^2 + 3.91925794\,\lambda - 827.0236$
Aug 2018: For $\lambda > 467$ nm, DPA = $-0.00496338\,\lambda^2 + 4.27806721\,\lambda - 915.5104$

* The standard deviation in PA determination is larger than usual for 2018JUL, this is mainly due to the measurement for HD 203532 being 2.8 degrees greater than the determined mean.

**Supplementary Table 4 | Linear Polarization Observations of Spica.**

| Modified Julian Date[a] | Orbital Phase | Telescope[b]/ Instrument | Run | Fil[c] | $\lambda_{eff}$[d] (nm) | Eff.[e] | Q/I (ppm) | U/I (ppm) |
|---|---|---|---|---|---|---|---|---|
| 57166.46191 | 0.23704 | AAT/HIPPI | 2015MAY | g′ | 462 | 0.887 | −172.0 ± 1.8 | 24.8 ± 1.8 |
| 57202.43786 | 0.19834 | AAT/HIPPI | 2015JUN | g′ | 462 | 0.887 | −205.8 ± 2.0 | 21.2 ± 1.9 |
| 57202.52971 | 0.22122 | AAT/HIPPI | 2015JUN | g′ | 463 | 0.890 | −206.9 ± 2.6 | 51.4 ± 2.8 |
| 57445.78081 | 0.81299 | AAT/HIPPI | 2016FEB | g′ | 462 | 0.859 | −179.4 ± 2.7 | 33.5 ± 2.7 |
| 57447.79340 | 0.31430 | AAT/HIPPI | 2016FEB | g′ | 462 | 0.860 | −139.0 ± 2.4 | −24.1 ± 2.5 |
| 57448.79605 | 0.56406 | AAT/HIPPI | 2016FEB | g′ | 462 | 0.860 | −300.5 ± 2.9 | −44.2 ± 2.9 |
| 57564.32900 | 0.34233 | AAT/HIPPI | 2016FEB | g′ | 462 | 0.860 | −113.8 ± 5.7 | −41.0 ± 6.1 |
| 57525.50647 | 0.67197 | UNSW/MHIPPI | M2016MAY | Cl | 451 | 0.741 | −283.5 ± 10.8 | 66.0 ± 10.8 |
| 57525.58860 | 0.69242 | UNSW/MHIPPI | M2016MAY | Cl | 453 | 0.748 | −265.5 ± 11.7 | 68.9 ± 11.4 |
| 57527.50765 | 0.17045 | UNSW/MHIPPI | M2016MAY | Cl | 451 | 0.741 | −212.8 ± 12.3 | 40.6 ± 12.3 |
| 57549.52778 | 0.65547 | UNSW/MHIPPI | M2016JUN | Cl | 453 | 0.748 | −265.8 ± 12.4 | 24.2 ± 12.4 |
| 57549.62670 | 0.68011 | UNSW/MHIPPI | M2016JUN | Cl | 465 | 0.794 | −226.5 ± 13.4 | 27.0 ± 14.1 |
| 57551.58819 | 0.16870 | UNSW/MHIPPI | M2016JUN | Cl | 458 | 0.770 | −177.0 ± 12.3 | 29.5 ± 12.1 |
| 57552.45161 | 0.38377 | UNSW/MHIPPI | M2016JUN | Cl | 451 | 0.741 | −4.1 ± 11.1 | 46.2 ± 10.3 |
| 57553.43005 | 0.62749 | UNSW/MHIPPI | M2016JUN | Cl | 451 | 0.741 | −323.8 ± 10.7 | −3.7 ± 10.7 |
| 57533.45623 | 0.63401 | UNSW/MHIPPI | M2016JUN | Cl | 451 | 0.741 | −321.4 ± 11.4 | 34.7 ± 10.9 |
| 57554.42968 | 0.87649 | UNSW/MHIPPI | M2016JUN | Cl | 451 | 0.741 | −191.6 ± 11.1 | −25.4 ±10.4 |
| 57555.42920 | 0.12546 | UNSW/MHIPPI | M2016JUN | Cl | 451 | 0.741 | −248.1 ± 10.3 | 57.9 ± 10.2 |
| 57568.41542 | 0.36022 | UNSW/MHIPPI | M2016JUL | Cl | 451 | 0.741 | −202.5 ± 10.4 | −38.2 ± 10.7 |
| 57568.55855 | 0.39587 | UNSW/MHIPPI | M2016JUL | Cl | 461 | 0.781 | −212.1 ± 12.5 | −82.5 ± 12.1 |
| 57570.43872 | 0.86420 | UNSW/MHIPPI | M2016JUL | Cl | 452 | 0.745 | −140.6 ± 10.3 | −15.7 ± 10.8 |
| 57583.39922 | 0.09256 | UNSW/MHIPPI | M2016JUL | Cl | 452 | 0.745 | -242.8 ± 18.2 | 65.0 ± 15.4 |
| 57586.43275 | 0.84818 | UNSW/MHIPPI | M2016JUL | Cl | 453 | 0.751 | −159.5 ± 10.6 | 5.3 ± 10.5 |
| 57592.37949 | 0.32947 | UNSW/MHIPPI | M2016JUL | Cl | 452 | 0.745 | −192.3 ± 22.9 | 21.6 ± 21.5 |
| 57592.39991 | 0.33455 | UNSW/MHIPPI | M2016JUL | Cl | 453 | 0.748 | −207.6 ± 17.9 | −21.7 ± 19.5 |
| 57592.42179 | 0.34000 | UNSW/MHIPPI | M2016JUL | Cl | 453 | 0.751 | −221.3 ± 15.5 | −26.0 ± 12.6 |
| 57594.38092 | 0.82800 | UNSW/MHIPPI | M2016JUL | Cl | 452 | 0.745 | −154.1 ± 11.1 | 11.9 ± 11.0 |
| 57595.37046 | 0.07449 | UNSW/MHIPPI | M2016JUL | Cl | 452 | 0.745 | −273.7 ± 12.0 | 60.3 ± 10.4 |
| 57595.39123 | 0.07966 | UNSW/MHIPPI | M2016JUL | Cl | 453 | 0.748 | −267.0 ± 10.1 | 7.0 ± 10.7 |
| 57608.37321 | 0.31336 | UNSW/MHIPPI | M2016JUL | Cl | 453 | 0.751 | −181.0 ± 21.5 | −12.0 ± 18.6 |
| 57611.39051 | 0.06495 | UNSW/MHIPPI | M2016JUL | Cl | 455 | 0.758 | −352.3 ± 10.9 | 26.7 ± 10.3 |
| 57613.36112 | 0.55581 | UNSW/MHIPPI | M2016JUL | Cl | 453 | 0.751 | −295.9 ± 10.7 | −21.7 ± 10.2 |
| 57614.40666 | 0.81624 | UNSW/MHIPPI | M2016JUL | Cl | 458 | 0.770 | −178.2 ± 11.2 | 14.9 ± 10.5 |
| 57615.38367 | 0.05961 | UNSW/MHIPPI | M2016JUL | Cl | 456 | 0.761 | −292.5 ± 10.5 | −34.5 ± 10.9 |
| 57616.39161 | 0.31068 | UNSW/MHIPPI | M2016JUL | Cl | 457 | 0.764 | −161.8 ± 11.9 | 73.7 ± 11.8 |
| 57627.36589 | 0.04428 | UNSW/MHIPPI | M2016JUL | Cl | 457 | 0.767 | −268.0 ± 11.6 | −12.8 ± 11.0 |
| 57627.38814 | 0.04982 | UNSW/MHIPPI | M2016JUL | Cl | 461 | 0.778 | −252.9 ± 11.6 | 17.9 ± 12.6 |
| 57631.36631 | 0.04075 | UNSW/MHIPPI | M2016JUL | Cl | 459 | 0.773 | −297.7 ± 12.4 | 34.9 ± 11.5 |
| 57631.38739 | 0.04600 | UNSW/MHIPPI | M2016JUL | Cl | 463 | 0.786 | −297.3 ± 13.1 | 24.4 ± 13.0 |
| 57647.36959 | 0.02703 | UNSW/MHIPPI | M2016JUL | Cl | 474 | 0.821 | −357.3 ± 51.2 | 41.9 ± 50.9 |
| 57745.73181 | 0.52822 | UNSW/MHIPPI | M2016OCT | Cl | 456 | 0.761 | −220.7 ± 14.0 | −223.4 ± 14.3 |
| 57747.72490 | 0.02468 | UNSW/MHIPPI | M2016OCT | Cl | 456 | 0.761 | −274.7 ± 12.8 | 32.4 ± 12.1 |
| 57769.69524 | 0.49730 | UNSW/MHIPPI | M2016OCT | Cl | 453 | 0.751 | −307.6 ± 13.7 | −42.4 ± 13.2 |
| 57780.64131 | 0.22387 | UNSW/MHIPPI | M2016OCT | Cl | 456 | 0.761 | −199.2 ± 12.0 | −13.4 ± 12.3 |
| 57788.66614 | 0.22279 | UNSW/MHIPPI | M2016OCT | Cl | 453 | 0.748 | −153.8 ± 11.1 | 56.7 ± 10.6 |
| 57804.59949 | 0.19165 | UNSW/MHIPPI | M2016OCT | Cl | 454 | 0.754 | −196.5 ± 10.6 | 48.6 ± 10.6 |
| 57853.51894 | 0.37707 | UNSW/MHIPPI | M2016OCT | Cl | 452 | 0.745 | −211.4 ± 13.3 | −36.8 ± 12.1 |
| 57857.53274 | 0.37687 | UNSW/MHIPPI | M2016OCT | Cl | 451 | 0.741 | −192.2 ± 10.6 | −76.9 ± 10.5 |
| 57871.43060 | 0.83871 | UNSW/MHIPPI | M2016OCT | Cl | 453 | 0.751 | −192.8 ± 10.8 | 5.0 ± 10.3 |
| 57871.45456 | 0.84468 | UNSW/MHIPPI | M2016OCT | Cl | 453 | 0.748 | −170.2 ± 10.4 | −20.8 ± 10.1 |
| 57871.62367 | 0.88747 | UNSW/MHIPPI | M2016OCT | Cl | 452 | 0.745 | −151.3 ± 10.1 | −65.7 ± 10.1 |
| 57871.64560 | 0.89227 | UNSW/MHIPPI | M2016OCT | Cl | 453 | 0.748 | −151.0 ± 10.4 | −81.5 ± 10.9 |
| 57873.43497 | 0.33798 | UNSW/MHIPPI | M2016OCT | Cl | 453 | 0.748 | −120.8 ± 10.3 | −104.0 ± 10.6 |
| 57873.45557 | 0.34311 | UNSW/MHIPPI | M2016OCT | Cl | 452 | 0.745 | −203.5 ± 10.4 | −62.8 ± 10.6 |
| 57873.47763 | 0.34861 | UNSW/MHIPPI | M2016OCT | Cl | 452 | 0.745 | −151.8 ± 10.8 | −49.1 ± 10.4 |
| 57878.35289 | 0.56299 | UNSW/MHIPPI | M2016OCT | Cl | 458 | 0.770 | −234.6 ± 11.5 | −50.7 ± 11.6 |
| 57878.37269 | 0.56793 | UNSW/MHIPPI | M2016OCT | Cl | 456 | 0.761 | −259.8 ± 11.5 | −47.8 ± 10.3 |
| 57880.41484 | 0.07661 | UNSW/MHIPPI | M2016OCT | Cl | 453 | 0.748 | −307.6 ± 10.6 | 33.0 ± 11.5 |
| 57880.43407 | 0.08140 | UNSW/MHIPPI | M2016OCT | Cl | 452 | 0.745 | −278.7 ± 11.1 | 67.1 ± 11.4 |
| 57881.34954 | 0.30943 | UNSW/MHIPPI | M2016OCT | Cl | 457 | 0.767 | −184.3 ± 11.7 | −13.5 ± 11.9 |
| 57881.36751 | 0.31391 | UNSW/MHIPPI | M2016OCT | Cl | 456 | 0.761 | −183.7 ± 11.5 | −40.0 ± 11.0 |
| 57881.39578 | 0.32095 | UNSW/MHIPPI | M2016OCT | Cl | 453 | 0.751 | −127.7 ± 11.1 | −29.7 ± 11.2 |
| 57883.60520 | 0.87130 | UNSW/MHIPPI | M2016OCT | Cl | 453 | 0.748 | −151.9 ± 10.6 | −57.0 ± 10.5 |
| 57883.62388 | 0.87595 | UNSW/MHIPPI | M2016OCT | Cl | 453 | 0.751 | −167.1 ± 10.7 | −38.9 ± 10.9 |
| 57888.45664 | 0.07975 | UNSW/MHIPPI | M2016OCT | Cl | 451 | 0.741 | −239.1 ± 11.9 | −3.6 ± 11.5 |
| 57888.53652 | 0.09965 | UNSW/MHIPPI | M2016OCT | Cl | 451 | 0.741 | −248.3 ± 11.3 | 96.3 ± 11.4 |
| 57894.41463 | 0.56384 | UNSW/MHIPPI | M2016OCT | Cl | 452 | 0.745 | −359.0 ± 11.2 | −50.4 ± 11.2 |
| 57894.43238 | 0.56826 | UNSW/MHIPPI | M2016OCT | Cl | 451 | 0.741 | −196.0 ± 11.6 | −38.4 ± 11.4 |
| 57896.37331 | 0.05173 | UNSW/MHIPPI | M2016OCT | Cl | 453 | 0.748 | −299.0 ± 11.7 | −55.4 ± 11.1 |
| 57898.34524 | 0.54292 | UNSW/MHIPPI | M2016OCT | Cl | 454 | 0.754 | −266.5 ± 10.0 | −35.9 ± 11.0 |
| 57898.36339 | 0.54744 | UNSW/MHIPPI | M2016OCT | Cl | 453 | 0.748 | −269.2 ± 10.9 | −31.2 ± 10.6 |
| 57898.56452 | 0.59754 | UNSW/MHIPPI | M2016OCT | Cl | 453 | 0.748 | −249.9 ± 11.0 | 12.3 ± 11.6 |
| 57898.58278 | 0.60209 | UNSW/MHIPPI | M2016OCT | Cl | 453 | 0.751 | −262.5 ± 11.4 | 28.3 ± 11.7 |
| 57928.33420 | 0.01291 | AAT/HIPPI | 2017JUN | Cl | 454 | 0.754 | −275.2 ± 3.1 | −73.3 ± 2.6 |
| 57929.33512 | 0.26223 | AAT/HIPPI | 2017JUN | Cl | 454 | 0.754 | −145.4 ± 3.6 | 13.7 ± 5.0 |
| 57930.33052 | 0.51020 | AAT/HIPPI | 2017JUN | Cl | 454 | 0.754 | −281.4 ± 2.2 | −83.0 ± 2.1 |
| 57933.34190 | 0.26029 | AAT/HIPPI | 2017JUN | Cl | 454 | 0.754 | −156.7 ± 3.5 | −16.0 ± 3.6 |
| 57934.33678 | 0.50810 | AAT/HIPPI | 2017JUN | Cl | 454 | 0.754 | −236.5 ± 2.1 | −85.5 ± 2.1 |
| 57935.33618 | 0.75704 | AAT/HIPPI | 2017JUN | Cl | 454 | 0.754 | −173.1 ± 2.3 | 26.5 ± 2.3 |
| 57936.33393 | 0.00558 | AAT/HIPPI | 2017JUN | Cl | 454 | 0.754 | −284.7 ± 1.9 | −94.4 ± 1.9 |
| 57977.34096 | 0.22007 | AAT/HIPPI | 2017AUG | Cl | 455 | 0.757 | −132.1 ± 1.9 | 32.3 ± 2.0 |
| 57978.34354 | 0.46981 | AAT/HIPPI | 2017AUG | Cl | 455 | 0.759 | −270.0 ± 2.0 | −111.7 ± 1.9 |
| 57980.34119 | 0.96741 | AAT/HIPPI | 2017AUG | Cl | 455 | 0.759 | −231.0 ± 2.2 | −75.4 ± 2.3 |
| 57981.34299 | 0.21695 | AAT/HIPPI | 2017AUG | Cl | 455 | 0.759 | −167.0 ± 2.3 | 27.3 ± 3.3 |
| 57984.34479 | 0.96467 | AAT/HIPPI | 2017AUG | Cl | 456 | 0.761 | −218.4 ± 3.5 | −106.5 ± 3.4 |
| 57985.34279 | 0.21326 | AAT/HIPPI | 2017AUG | Cl | 456 | 0.761 | −177.0 ± 3.1 | 55.3 ± 2.8 |
| 58152.69840 | 0.90011 | AAT*/HIPPI2 | 2018FEBB | Cl | 454 | 0.686 | −180.1 ± 2.0 | −59.2 ± 1.9 |
| 58153.75202 | 0.16256 | AAT*/HIPPI2 | 2018FEBD | Cl | 453 | 0.683 | −219.7 ± 1.8 | 75.9 ± 1.8 |
| 58154.77591 | 0.41760 | AAT*/HIPPI2 | 2018FEBD | Cl | 453 | 0.683 | −218.5 ± 3.2 | −117.8 ± 2.6 |
| 58200.79225 | 0.87988 | AAT/HIPPI2 | 2018MAR | Cl | 462 | 0.726 | −157.6 ± 1.7 | −49.8 ± 1.6 |
| 58201.80348 | 0.13177 | AAT/HIPPI2 | 2018MAR | Cl | 463 | 0.731 | −231.0 ± 1.8 | 24.8 ± 2.5 |
| 58203.79980 | 0.62904 | AAT/HIPPI2 | 2018MAR | Cl | 463 | 0.731 | −260.4 ± 1.9 | −22.3 ± 1.8 |
| 58204.79716 | 0.87747 | AAT/HIPPI2 | 2018MAR | Cl | 463 | 0.731 | −188.4 ± 1.8 | −59.8 ± 1.7 |
| 58207.79703 | 0.62471 | AAT/HIPPI2 | 2018MAR | Cl | 463 | 0.733 | −276.3 ± 1.8 | 12.8 ± 1.9 |
| 58210.53357 | 0.30636 | AAT/HIPPI2 | 2018MAR | Cl | 459 | 0.717 | −176.4 ± 4.2 | −62.7 ± 4.9 |
| 58210.57468 | 0.31660 | AAT/HIPPI2 | 2018MAR | Cl | 459 | 0.714 | −172.8 ± 4.2 | −73.3 ± 4.5 |
| 58212.77341 | 0.86429 | AAT/HIPPI2 | 2018MAR | Cl | 463 | 0.731 | −184.9 ± 1.9 | −58.3 ± 2.0 |
| 58242.62498 | 0.30006 | WSU/HIPPI2 | 2018MAY | Cl | 461 | 0.723 | −153.2 ± 9.6 | −59.1 ± 10.0 |
| 58242.64011 | 0.30383 | WSU/HIPPI2 | 2018MAY | Cl | 461 | 0.725 | −154.1 ± 10.5 | −54.9 ± 10.9 |
| 58242.65395 | 0.30727 | WSU/HIPPI2 | 2018MAY | Cl | 462 | 0.728 | −196.4 ± 11.1 | −77.9 ± 8.9 |
| 58242.66768 | 0.31069 | WSU/HIPPI2 | 2018MAY | Cl | 463 | 0.731 | −161.2 ± 11.2 | −60.0 ± 11.3 |
| 58242.68237 | 0.31435 | WSU/HIPPI2 | 2018MAY | Cl | 463 | 0.733 | −189.2 ± 10.9 | −52.3 ± 11.3 |
| 58247.54748 | 0.52621 | WSU/HIPPI2 | 2018MAY | Cl | 460 | 0.717 | −315.0 ± 9.8 | −37.3 ± 10.4 |
| 58249.52182 | 0.01800 | WSU/HIPPI2 | 2018MAY | Cl | 460 | 0.717 | −312.4 ± 11.7 | −38.0 ± 10.7 |
| 58249.53773 | 0.02196 | WSU/HIPPI2 | 2018MAY | Cl | 460 | 0.717 | −296.3 ± 11.8 | −48.0 ± 13.7 |
| 58284.50760 | 0.73266 | UNSW/MHIPPI | M2018JAN | Cl | 453 | 0.748 | −278.5 ± 20.5 | −5.5 ± 19.3 |
| 58311.33394 | 0.41487 | AAT/HIPPI2 | 2018JUL | Cl | 459 | 0.703 | −210.8 ± 2.1 | −81.3 ± 1.9 |
| 58316.33830 | 0.66142 | AAT/HIPPI2 | 2018JUL | Cl | 459 | 0.703 | −243.9 ± 1.8 | 30.6 ± 1.9 |
| 58328.40509 | 0.66715 | UNSW/MHIPPI | M2018JAN | Cl | 453 | 0.751 | −226.7 ± 13.2 | 43.9 ± 14.3 |
| 58334.40236 | 0.16102 | UNSW/MHIPPI | M2018JAN | Cl | 454 | 0.754 | −218.2 ± 11.3 | 72.7 ± 11.7 |
| 58335.35876 | 0.39925 | UNSW/MHIPPI | M2018JAN | Cl | 453 | 0.748 | −237.1 ± 20.4 | −88.0 ± 24.0 |
| 58335.43132 | 0.41732 | UNSW/MHIPPI | M2018JAN | Cl | 457 | 0.767 | −206.1 ± 14.0 | −40.7 ± 13.9 |
| 58343.38059 | 0.39742 | UNSW/MHIPPI | M2018JAN | Cl | 455 | 0.758 | −229.3 ± 13.5 | −58.9 ± 13.0 |
| 58346.37939 | 0.14439 | AAT/HIPPI2 | 2018AUG | Cl | 462 | 0.626 | −193.0 ± 2.4 | 63.2 ± 2.1 |
| 58347.38759 | 0.39553 | AAT/HIPPI2 | 2018AUG | Cl | 462 | 0.629 | −207.1 ± 2.0 | −118.5 ± 2.3 |
| 58348.36094 | 0.63798 | AAT/HIPPI2 | 2018AUG | Cl | 461 | 0.620 | −268.6 ± 4.7 | 64.2 ± 4.6 |
| 58350.35545 | 0.13480 | AAT/HIPPI2 | 2018AUG | Cl | 461 | 0.620 | −203.1 ± 2.0 | 74.2 ± 1.9 |
| 58351.39224 | 0.39305 | AAT/HIPPI2 | 2018AUG | Cl | 463 | 0.635 | −182.0 ± 1.9 | −87.0 ± 2.2 |
| 58352.39406 | 0.64260 | AAT/HIPPI2 | 2018AUG | Cl | 464 | 0.638 | −258.4 ± 2.1 | 42.3 ± 2.1 |
| 58357.38917 | 0.88684 | AAT/HIPPI2 | 2018AUG | Cl | 465 | 0.582 | −170.1 ± 3.2 | −58.4 ± 2.7 |
| 58359.37215 | 0.38078 | AAT/HIPPI2 | 2018AUG | Cl | 463 | 0.535 | −193.0 ± 2.5 | −91.7 ± 2.5 |
| 58362.36071 | 0.12520 | AAT/HIPPI2 | 2018AUG | Cl | 463 | 0.535 | −194.1 ± 2.7 | 56.7 ± 2.7 |
| 58363.36297 | 0.37486 | AAT/HIPPI2 | 2018AUG | Cl | 464 | 0.538 | −176.5 ± 2.7 | −79.3 ± 2.6 |

**Notes:**

[a] Heliocentric modified Julian Date (= JD – 2400000.5) of mid point of observation.
[b] Telescopes are: UNSW, 35 cm Schmidt Cassegrain telescope at UNSW Observatory, Sydney; WSU, Western Sydney University's 60 cm Ritchey Chretien Telescope at Penrith Observatory; AAT, 3.9 m Anglo-Australian Telescope.
[c] Cl denoting clear (no filter) or SDSS g' filter.
[d] The effective wavelength of the observation in nm.
[e] The efficiency correction – the raw value is divided by this number to give the tabulated values.
* Indicates AAT f/15, all other AAT observations are f/8.